\documentclass[aps,prl,twocolumn,showpacs,groupedaddress]{revtex4}
\usepackage{graphicx}

\begin{document}

\title{High-power
supercontinuum generation in dielectric-coated metallic hollow waveguides}
\author{A. Husakou}
\email{[]gusakov@mbi-berlin.de}
\affiliation{Max Born Institute of Nonlinear Optics and Short Pulse Spectroscopy, Max
Born Str 2a, D-12489 Berlin, Germany}
\author{J. Herrmann}
\affiliation{Max Born Institute of Nonlinear Optics and Short Pulse Spectroscopy, Max
Born Str 2a, D-12489 Berlin, Germany}
\date{\today}

\begin{abstract}
In this Letter we theoretically study a novel approach for soliton-induced
supercontinuum generation based on the application of metallic
dielectric-coated hollow waveguides. The low loss of such waveguides permits the
use of smaller diameters with enhanced dispersion control and enables the
generation of two-octave-broad spectra with unprecedentedly high spectral
peak power densities up to five orders of magnitude larger than in standard
PCFs with high coherence. We also predict that high-power supercontinua in
the vacuum ultraviolet can be generated in such waveguides.
\end{abstract}

\maketitle








It is well known that lasers generate strongly directed coherent light with
the highest possible brightness. However, a typical laser is a
quasi-monochromatic source emitting light of only one color. Many
applications require light sources which share with a laser its
unidirectional and coherent properties but span the whole spectral range of
a rainbow like the sun or an electric bulb. The latter sources, however, are
unable to emit coherent, unidirectional and bright radiation.

A milestone on the way towards a coherent white-light high-brightness source
(supercontinuum, SC) was achieved by the application of photonic crystal fibres
(PCF) \cite{1,2}. When a femtosecond pulse with only nJ energy from a laser
oscillator is focused into such fiber, a dramatic conversion from narrow band
light to two-octaves-broad spectra was observed \cite{3}. The discovery of SC
generation in PCFs encouraged extensive research activities (see e.g. Refs.%
\cite{4,12,13,14,11,15,16,26}) and is now considered to be one of the leading-edge
research areas in photonics. This coherent
white-light source is applied in physics, chemistry, biology,
telecommunication and medicine using numerous methods ranging from the
frequency comb method \cite{5,6}, optical coherence tomography \cite{7}, 
absorption spectroscopy\cite{as}, and flow cytometry to WDM in telecommunication and
others. There are now a few commercial suppliers for these purposes 
(see e.g. \cite{imam}). 

Supercontinuum generation in PCFs is based on an effect for spectral
broadening \cite{4} which differ from the main previously known mechanism
for spectral broadening by self-phase modulation. It is connected with
the soliton dynamics in the anomalous dispersion region of the PCFs. The
high refractive-index contrast in PCFs leads to a shift of the zero dispersion wavelength to
the visible and anomalous dispersion at the wavelength of typical lasers, such as
 a Ti:sapphire laser. Therefore the input pulse splits into several
fundamental solitons with different amplitudes and distinct red-shifted
input frequencies. Per definition, a fundamental soliton does not change its
energy, however if it is perturbed by third- and higher-order dispersion
of the fibre it can emit blue-shifted nonsolitonic radiation (also known as
dispersive radiation) at a wavelength determined by phase matching \cite{17}.
Since multiple solitons with different central frequencies emit radiation at
distinct frequency intervals, a very large spectral range is covered.
Subsequent experimental studies \cite{12,13,14} by many groups provided
evidence for this soliton-induced mechanism for SC generation.

\begin{figure}[!b]
\includegraphics[width=0.35\textwidth]{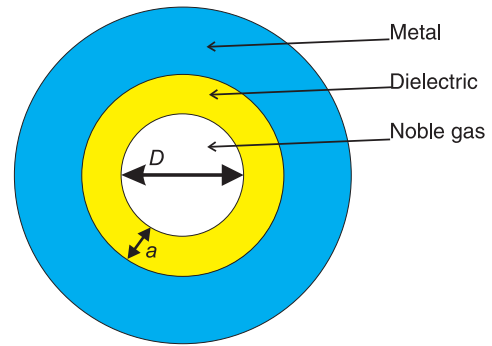}
\caption{Scheme of the hollow waveguide for supercontinuum generation. 
 }
\label{HE_sch}
\end{figure}

\begin{figure*}[t]
\includegraphics[width=0.75\textwidth]{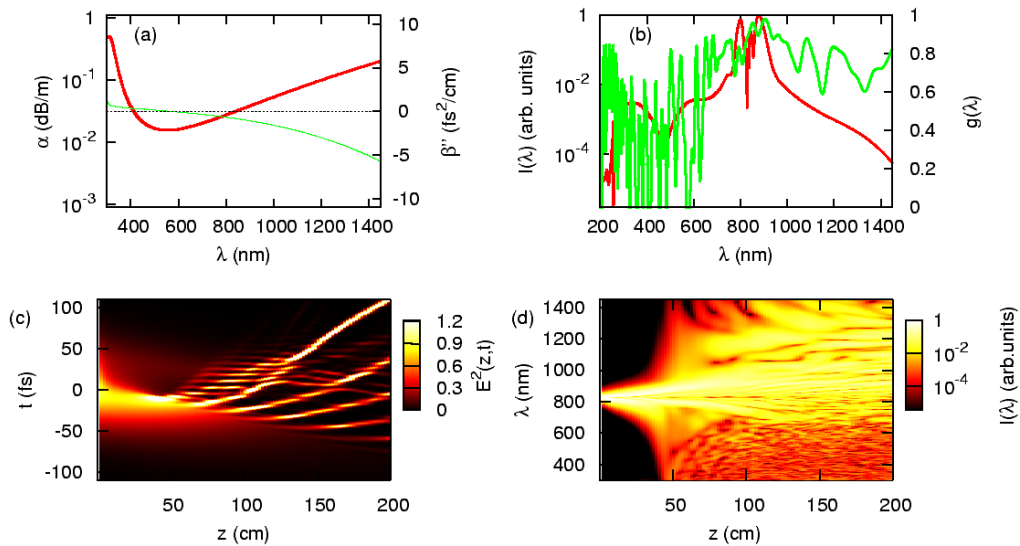}
\caption{High-power supercontinuum generation. Waveguide characteristics
(a), generated output supercontinuum (red) and first- order coherence
(green) (b), and evolution of temporal shape (c) and spectrum (d) in a $D$ =
80 $\protect\mu$m silver waveguide coated with a 45-nm layer of fused silica
and filled with argon at 1 atm. The input 100-fs pulse at 800 nm has the
intensity of 50 TW/cm$^2$. In (a), loss (red) and group velocity dispersion
(green) are presented. The roughness size $\protect\sigma$ equals 100 nm.
The spectrum in (b) is presented after 50-cm propagation. }
\label{HE}
\end{figure*}

Supercontinuum spectra from PCFs are 10$^{6}$ times brighter than sunlight ($%
\sim $10$^{3}$ W/cm$^{2}$/sterad) and have the same bandwidth.
Unfortunately, the small radii in PCFs and material damage at few TW/cm$^2$ limit severely
the maximum peak power densities to tens of W/nm. In this Letter we propose
and study a novel approach for supercontinuum generation in metallic hollow
waveguides coated with a dielectric. We predict that in such waveguides
two-octave broad SC with up to five orders of magnitude higher spectral peak
power densities than in PCFs can be generated. We also predict the
generation of UV/VUV supercontinua in the range from 160 to 540 nm from such type
of waveguide, a spectral range which can not be achieved by the previously
used methods but which is of particular importance in many applications. The
key idea in this approach is to use a hollow-core metal-dielectric waveguide filled with a
noble gas instead of the solid-core microstructure fibers to increase the
diameter to a range between 20 to 80$\mu $m and the damage threshold. The
introduction of such waveguides has the aim to fulfill the conditions of
small loss and anomalous dispersion at optical frequencies which are the
basic requirements for the soliton-induced mechanism of supercontiunuum
generation, similar to the situation in microstructure fibers \cite{4}. In recent years the
application of standard dielectric hollow waveguides has led to impressive
progress in ultrafast optics such as the generation of few-cycle \cite{9} and
attosecond pulses \cite{10} and high-order harmonics \cite{18}. However, these
waveguides can not be used for the purpose studied here,
because anomalous
dispersion at optical frequencies and high pressures can only be achieved for small
diameters in the range of 20-80$\mu $m, 
 for which the loss is too high. 

Dielectric-coated metallic hollow
waveguides can be produced by chemical vapor deposition \cite{19,20}. In Fig.
1, the geometry of such waveguide is presented. The hollow core of the fibre
with diameter $D$ is surrounded by a metallic cladding (blue) coated by a
dielectric material such as silica (yellow) with thickness a on the inner
side.

For the numerical simulations we use a generalized model based on the
propagation equation for forward-going waves \cite{4} with inclusion of
higher-order transverse modes and the effects of ionization and plasma
formation in the waveguide. The Fourier transform $\vec{E}(z,x,y,\omega )$
of the electromagnetic field $\vec{E}(z,x,y,t)$ in the waveguide can be
represented by $\vec{E}(z,x,y,\omega)=\sum_{j}E_{j}(z,\omega
)F_{j}(r,\omega)$, where $F_{j}(r,\omega)$ is the transverse mode profiles
of the $j$-th mode and $E_{j}(z,\omega)$ describes the evolution of field with propagation.
 The inclusion of higher-order transverse modes takes into
account a possible energy transfer to higher-order modes by the nonlinear
polarization. We consider only linearly-polarized fields and therefore only
EH$_{1j}$ modes with the same polarization. The
following first-order differential equation can be written for 
$E_{j}(z,\omega )$ in the EH$_{1j}$ mode \cite{4}: 
\begin{eqnarray}
\frac{\partial E_{j}(z,\omega )}{\partial z} &=&i\beta _{j}(\omega
)E_{j}(z,\omega )-\frac{\alpha _{j}}{2}(\omega )E_{j}(z,\omega )  \nonumber
\\
&&+\frac{i\omega ^{2}}{2c^{2}\epsilon _{0}\beta _{j}(\omega )}%
P_{NL}^{(j)}(z,\omega )
\end{eqnarray}%
where $\beta _{j}(\omega )$ and $\alpha _{j}(\omega )$ are the wavenumber
and the loss.
This equation neglects 
backward-propagating components and contains only
the first-order derivative over the propagation coordinate, but in
difference to the nonlinear Schr\"{o}dinger equation (NSE) it does not rely
on the slowly-varying envelope approximation and  refer to the field
components $E_{j}(z,\omega )$ and not to the amplitudes of the field. This
approach allows to model the evolution of fs pulses with extremly broad
spectra. The quantities $\beta_{j}(\omega )$, $\alpha_{j}(\omega )$ and $%
F_{j}(r,\omega )$ are calculated by the transfer-matrix approach assuming a
circular waveguide structure as shown in Fig. 1, including the bulk dispersion of argon.
Additionally, roughness loss is included in the calculation of $\alpha
_{j}(\omega )$ using the model of pointlike scatterers; for further details
of the transfer matrix theory and roughness loss calculation, see Ref. \cite%
{22}.

\begin{figure*}[!t]
\includegraphics[width=0.75\textwidth]{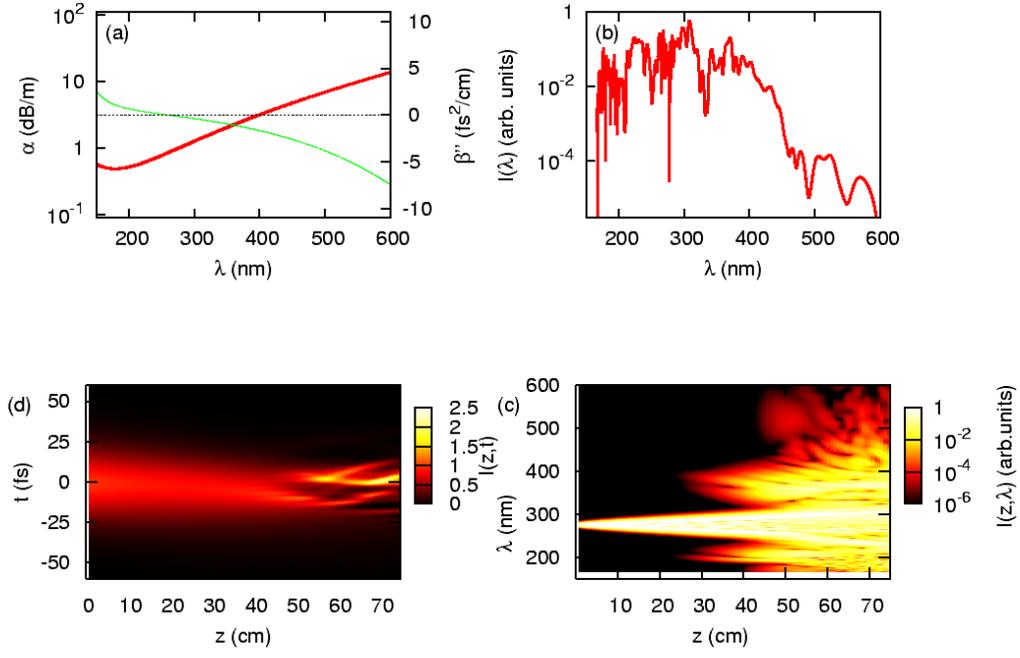}
\caption{UV/VUV supercontinuum generation. Waveguide characteristics (a),
generated output supercontinuum (b), and evolution of the temporal shape (c)
and spectrum (d) in a 20-$\protect\mu$m aluminum waveguide with a 10-nm
coating and filled with argon at 0.5 atm. The input 50-fs pulse at 266 nm
has the input intensity of 40 TW/cm$^2$. In (a), loss (red) and group
velocity dispersion (green) are presented. }
\label{UV}
\end{figure*}

The Fourier transform of the $j$th component of the nonlinear polarization 
 $P_{NL}^{(j)}(z,\omega )$ is given by 
\begin{equation}
P_{NL}^{(j)}(z,\omega )=\int_{0}^{R}\!\!2\pi rF_{j}(r,\omega)\int_{-\infty
}^{\infty }\!\!\exp (i\omega t)P_{NL}(z,r,t)drdt
\end{equation}
Here $P_{NL}(z,r,t)$ includes the Kerr nonlinearity,
the plasma-induced refraction index change and absorption of light
energy due to ionization. A theoretical description of light propagation in
gases in the presense of photoionization and plasma can be found in numerous
papers (see e.g. Ref. \cite{plas}). Taking these effects into account we can
write for $P_{NL}(z,r,t)$:   
\begin{eqnarray}
P_{NL}(z,r,t) &=&\epsilon _{0}\chi _{3}E^{3}(z,r,t)-\rho (z,r,t)ed(z,r,t)- 
\nonumber \\
&&{E_{g}\epsilon _{0}c}\int_{-\infty }^{t}\frac{E(z,r,t)}{\tilde{I}(z,r,t)}%
\frac{\partial \rho (z,r,t)}{\partial t}dt
\end{eqnarray}%
where $\chi _{3}=(4/3)c\epsilon _{0}n_{2}$ is the third-order polarizability
of the gas filling and $\tilde{I}(z,r,t)$ is the intensity averaged over few
optical periods. For argon we have $n_{2}=1\times 10^{-19}$ cm$^{2}$/W at 1
atm, $E_{g}$=15.95 eV is the ionization potential, $\rho (z,r,t)$ is the electron
density and $d(z,r,t)$ is the mean free-electron displacement in the
electric field. The evolution of the two latter quantities is described by 
\begin{equation}
\frac{\partial \rho (z,r,t)}{\partial t}=(N_{0}-\rho (z,r,t))\Gamma (\tilde{I
}(z,r,t))
\end{equation}
\begin{equation}
\frac{\partial ^{2}d(z,r,t)}{\partial t^{2}}=-eE(z,r,t)/m_{e}
\end{equation}
where $\Gamma (\tilde{I}(z,r,t))$ is the photoionization rate calculated
from the Faisal-Reiss-Keldysh model\cite{frk} which describes both the
multiphoton and tunnelling regime and $N_{0}$ is the initial argon density.
 The coherence properties of the supercontinua \cite{16,26}
are described by the first-order coherence function $g(\lambda )$ defined as 
\begin{equation}
g(\lambda )=\Re \left[ \frac{<E_{b}(\lambda )E_{a}^{\ast }(\lambda
)>_{ab,a\neq b}}{<E_{a}(\lambda )E_{a}^{\ast }(\lambda )>_{a}}\right] 
\end{equation}%
where the indices $a,b$ denote the different input realizations
and $E_{a}(\lambda)$ is the spectrum for the realisation $a$.
Pulse evolution with propagation is modelled for each realisation of the
input quantum shot noise, which is added to the input pulse. These
realisations are implemented as described in Ref. \cite{26}.  Equations (1)-(5) are solved by the split-step Fourier method with Runge-Kutta nonlinear steps.  

First we study the loss and dispersion properties of a silver waveguide with
a diameter of 80$\mu $m coated from the inner side with a \ 45nm layer from
fused silica. In Fig. 2(a) can be seen that the loss remains relatively low
in the range of 10$^{-2}$-10$^{-1}$ dB/m for wavelengths from 0.4 $\mu $m to
1.4 $\mu $m. This means that at most 10\% of the energy is lost during a
propagation over 1 m, while in a conventional dielectric hollow waveguide
with the same diameter, roughly 80\% of the energy is lost. The physical
reason for this difference is that the reflection coefficient for grazing
incidence of light is higher for fused-silica-coated silver than for a layer
of fused silica. The increased transmission of dielectric-coated metallic
waveguides allow a reduction of the diameter of the waveguides while keeping
the losses acceptable. This, in turn, leads to significant modification of
the dispersion properties and an extended range of anomalous dispersion\cite%
{22}. The group velocity dispersion, illustrated in Fig. 2(a) by the green
curve, is anomalous for $\lambda >570$ nm at 1 atm of the argon filling. 

The optical properties of the studied waveguide as illustrated in Fig. 2(a)
allows the conjecture that soliton-induced supercontinuum generation can
be generated in such waveguide. To study this question we consider the
evolution of a 100-fs input pulse at 830 nm with a peak intensity of 50 TW/cm%
$^{2}$ propagating a maximum distance of 2 m in the above-described fiber with 10 modes included in the simulation. We assume that the input field is 
completely in the EH$_{11}$ mode, but during propagation roughly 20\% of energy 
is transferred to higher-order modes.
It can be seen in Fig. 2(b) that the generated radiation cover the spectral
range from 250 nm to 1200 nm, with a total width corresponding to more than
two octaves. This supercontinuum is generated already after 50 cm of
propagation, as shown in Fig. 2(d). In Fig. 2(b) the first-order coherence
is shown by the green curve demonstrating a high average value of 0.92. The supercontinuum generation is stable against manufacturing imperfections and variation of input pulse parameters. The energy fraction in the cladding is around 10$^{-3}$, which guarantees that it is not damaged. The
wavelength-averaged peak power spectral density is about 10$^{6}$ W/nm with
an intensity at the output of about 10$^{14}$ W/cm$^{2}$. The obtained
spectral power density is 10$^{5}$ times higher than in standard PCFs. For
comparison, the Sun at its surface has a brightness of 10$^{3}$ W/cm$^{2}$%
/sterad, which is roughly 10$^{11}$ times lower than the results predicted
here. The temporal shape presented in Fig. 2(c) shows that the pulse is
split into many solitons and background radiation which move with different
velocities. The supercontinuum is explained by the same mechanism as in
photonic crystal fibres due to the emission of nonsolitonic radiation by
several frequency-shifted solitons\cite{4}.

Supercontinua from PCFs have the shortest wavelengths in the ultraviolet at
about 300 nm. There exists a strong interest in molecular physics and
material science in extending the achievable spectrum still further towards
the vacuum ultraviolet (VUV). Here we show that dielectric-coated metallic
hollow waveguides can also be used for this purpose and predict a high-power VUV
supercontinuum source, which is based on an appropriately
designed fused-silica-coated aluminium waveguides filled with argon.

We consider an optimized aluminum waveguide with the diameter of only 20 $\mu
$m coated by a 10-nm layer of fused-silica. The linear optical properties of
this waveguide are shown in Fig. 3(a), demonstrating moderate loss and
anomalous dispersion for $\lambda> 260$ nm at 1 atm. In Fig. 3(b)-(d) the
generation of an ultraviolet/vacuum ultraviolet supercontinuum from a third
harmonic pulse of a Ti:sapphire amplifier laser system with input pulses at
266 nm and intensity of 40 TW/cm$^2$ is illustrated. The evolution of
spectra (d) and the temporal shape (c) demonstrates an initial generation of
two side peaks by four-wave mixing followed by fission into several solitons
and soliton emission. The output spectrum reaches from 160 to 540 nm, as
shown by the red curve in Fig. 3(b).


In conclusion, we studied the generation of high-power supercontinua in
specially designed metallic dielectric coated waveguides. We predicted that
supercontinua with two-octave width, spectral power densities in the range
of MW/nm five orders of magnitude higher than in microstructured fibers, and
high coherence, can be achieved in such waveguides. Additionally, we predicted
the generation of UV/VUV supercontinuum in an aluminum hollow waveguide.

The findings that we report could have applications in a wide range of
fields. Let us remark just a few. In combination with a multichannel
frequency filter the high- power SC source can replace many lasers at
separated frequencies including wavelengths where lasers are not available
(as e.g. in the UV/VUV). The predicted VUV supercontinuum could lead to
advances in VUV frequency comb spectroscopy \cite{23}. Further on, direct
difference frequency mixing of two portions of the SC \cite{24} eliminates
the carrier-offset frequency and its noise which can be used in frequency
metrology. The detection sensitivity of nonlinear spectroscopic methods such
as CARS microscopy \cite{25} can be significantly increased by an increase of
the CS intensity.

\end{document}